\documentclass[conference]{IEEEtran}
\usepackage{graphicx}

\usepackage{cite}

\usepackage{amsmath}

\usepackage{cite}

\ifCLASSINFOpdf
\else
\fi
\begin{document}
%
\title{Wideband Signal Localization\\ with Spectral Segmentation}
%
%
%

\author{
    \IEEEauthorblockN{Author1\IEEEauthorrefmark{1}, Author2\IEEEauthorrefmark{2}, Author3\IEEEauthorrefmark{2}, Author4\IEEEauthorrefmark{1}}
    \IEEEauthorblockA{\IEEEauthorrefmark{1}Institution1
    \\\{1, 4\}@abc.com}
    \IEEEauthorblockA{\IEEEauthorrefmark{2}Institution2
    \\\{2, 3\}@def.com}
}
\author{
    \IEEEauthorblockN{Nathan~West\IEEEauthorrefmark{1}\IEEEauthorrefmark{2}
    Tamoghna~Roy\IEEEauthorrefmark{1},
    Tim~O'Shea\IEEEauthorrefmark{1}}

\vspace{1em}
\IEEEauthorblockA{\IEEEauthorrefmark{1}DeepSig Inc.\\
Arlington, VA\\
{first name@deepsig.io}}\\
\vspace{1em}
\IEEEauthorblockA{\IEEEauthorrefmark{2}Oklahoma State University\\
Stillwater, OK}
}

\markboth{IEEE Asilomar Conference on Signals, Systems, and Computers}%
{West \MakeLowercase{\textit{et al.}}: Wideband Signal Localization with Spectral Segmentation}
%



\maketitle

\begin{abstract}
Signal localization is a spectrum sensing problem that jointly detects the presence of a signal and estimates a center frequency and bandwidth. This is a step beyond most spectrum sensing work which estimates "present" or "not present" detections for either a single channel or fixed sized channels. We define the signal localization task, present the metrics of precision and recall, and establish baselines for traditional energy detection on this task. We introduce a new dataset that is useful for training neural networks to perform this task and show a training framework to train signal detectors to achieve the task and present precision and recall curves over SNR. This neural network based approach shows an 8 dB improvement in recall over the traditional energy detection approach with minor improvements in precision.
\end{abstract}


%
\IEEEpeerreviewmaketitle

\section{Introduction}
%
%
%
%
\IEEEPARstart{S}{ensing} the electromagnetic spectrum for the presence of signals is a well studied topic with defense, regulatory/policy, and industrial applications. The generic goal is to identify if a given portion of the spectrum is occupied by a signal. The exact application determines parameters of interest to estimate; although nearly every application requires or benefits from the signal bandwidth, center frequency, and modulation. Estimating the presence of a signal and these parameters is useful for physical security by knowing when wireless devices enter a physical area, policy by knowing how occupied spectrum is, license enforcement by recognizing interfering devices, and effectively using whitespaces without causing interference for the primary user. The general problem of spectrum sensing for these applications is actually a combination of two tasks:
\begin{enumerate}
    \item signal detection
    \item parameter estimation
\end{enumerate}

This distinction is important because although there are well established techniques for portions of these tasks with their own metrics there is relatively little research treating the problem of detecting wireless signal and identifying parameters such as frequency edges, start time, and duration as a joint problem. This combined task considers the case of a wideband receiver in which a signal can appear at any frequency, bandwidth, and time. We will call this wideband spectrum sensing because the sample bandwidth is much wider than an individual signal bandwidth (such that multiple signals may appear within the sample bandwidth).

The system model under consideration (shown in Equation \ref{eq:received_signal_model}) is that of a received sample stream $r(t)$ that is the sum of $N$ signals, $s_n(t)$ that each pass through an independent channel and additive white gaussian noise (AWGN), $N_0(t)$ generated at the receiver.

\begin{equation}
	\label{eq:received_signal_model}
	r(t) = \sum_{n=1}^N C_n(s_n(t)) + N_0(t)
\end{equation}

\subsection{Localizing Signals in Time and Frequency}

This is distinct from most of the spectrum sensing work that focuses on an individual channel to make a binary "present" or "not present" decision such as \cite{snr-walls,wb-nb-detection,snr-walls-features,energy-detection} because a decision must be made for a unique region within time-frequency space as well as an accurate prediction of that time-frequency space. Some works have relaxed that constraint to multiple channels where the channels have defined widths and centers \cite{adaptive-threshold}. Finally, some work has been done with no assumption of channel width and center frequencies, but simply to estimate spectrum usage without concern for identifying accurate estimations of decisions \cite{xgcomms}. A comprehensive survey of many of these techniques is given in \cite{survey-of-spectrum-sensing}. This problem of identifying where in time and frequency space a signal may appear without restriction on channel placement and width will be called signal localization.

The most similar work to signal localization is the localization algorithm with double thresholding (LAD) \cite{lads,lad-acc,localization-fcme} which jointly detects signals with an estimate on their upper and lower frequency bounds (equivalent to a center frequency and bandwidth estimate). This was extended to also estimate start and stop time boundaries to give LAD-2D in \cite{lad2d}. That work presents experimental results using a QPSK signal in AWGN. The results presented show the fraction of times the algorithm detected the correct number of signals (1) over a range of SNRs. The LAD algorithm has above 90\% correct number of detections above -1.8 dB. The LAD with adjacent cluster combining shows 84\% correct number of detections at -8 dB and 100\% correct at -6 dB and higher. LAD-2D shows 90\% correct at -8 dB and 100\% correct at -7 dB and higher SNR.

\subsection{Approaches}

\begin{figure}
    \centering
    \includegraphics[width=\columnwidth]{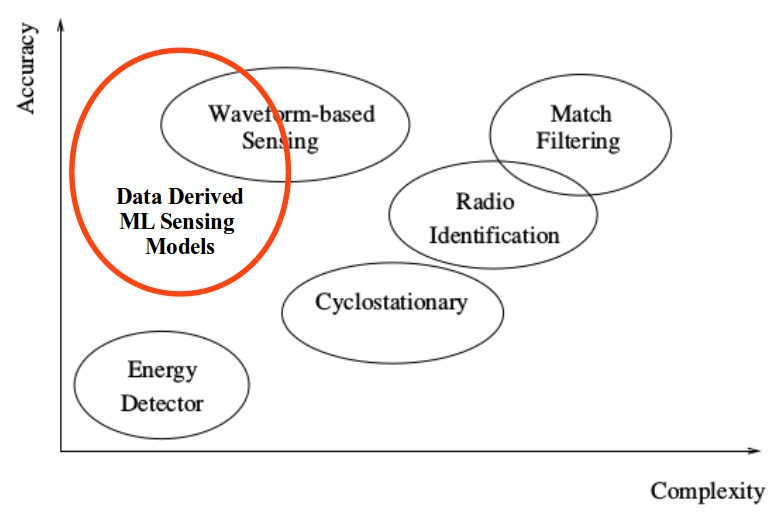}
    \caption{A trade-space of spectrum sensing algorithms. Traditional algorithms tend to be more complex to achieve higher accuracy. We show that a relatively simple neural network-based approach can achieve high accuracy with low complexity.}
    \label{fig:tradespace}
\end{figure}

Traditionally, this type of spectrum sensing is approached with a radiometer or using cyclostationary approaches. Figure \ref{fig:tradespace} shows a trade space of these approaches with our perception of accuracy and complexity. Cyclostationary approaches such as \cite{mathys-multi-sc} are challenging to use when a wide variety of signals may appear for an unknown time duration. We will compare results using spectrum sensing with the channelized radiometer and a neural-network based approach which can achieve far greater accuracy.

\section{Metrics}

\begin{figure}[!t]
\centering
\includegraphics[width=\columnwidth]{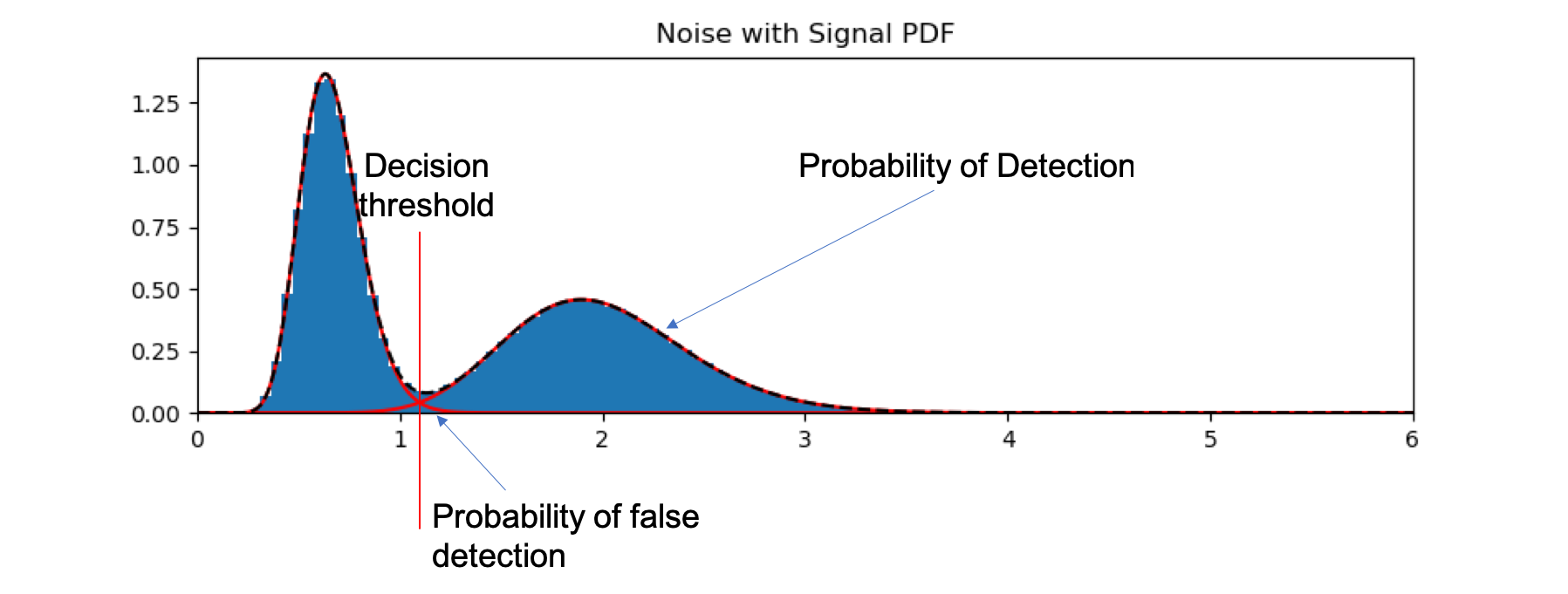}
\caption{Histogram showing two overlapping distributions summed together and the quantities that make up $P(fa)$ and $P(D)$. The noise distribution of a detector is at left and the distribution of the signal of interest is at right. The $P(fa)$ can only be known if the noise distribution on the detector output can be approximated. The $P(D)$ can only be known is the signal distribution can be approximated.}
\label{fig:histogram-pfa-pd}
\end{figure}

The binary spectrum sensing problem frequently uses measures of probability of false alarm ($P(fa)$) and probability of correct detection ($P(D)$). Often these values will be shown in a receiver operating characteristic curve that shows these values changing with SNR. These are valid metrics when the distribution of the output of a detector can be measured (which allows knowing $P(fa)$ and when the distribution of the signal of interest can be approximated (which allows knowing $P(D)$. Figure \ref{fig:histogram-pfa-pd} shows a histogram of a noise and signal of interest distribution to visualize the importance of knowing these distributions before $P(fa)$ and $P(D)$ can be computed. For binary decision making of "present" or "not present" within a define channel it is sufficient to assume a gaussian or rayleigh approximation in signal power and either use AWGN or model the detector output using an appropriate random variable transformations using AWGN in. However, in the signal localization task there is a decision on present or not present being made jointly with a regression on channel bandwidth and center frequency. This implies many joint decisions being made which overwhelms the feasibility of approximating the regression outputs. Since the signal of interest is unconstrained in center frequency and bandwidth (except in this case that we are limited in bandwidth by the sample bandwidth) knowing the $P(D)$ is limited by estimating the underlying probability of a given signal, such as 2 MHz QPSK at a given center frequency. Since this is not a knowable distribution, new metrics are required.

Rather than attempting an estimation of the underlying probabilities, we propose the measurement based metrics of precision and recall from the field of information retrieval. These have been combined with the Jaccard index (also known as the intersection over union or IoU) in the image domain's field of object localization \cite{pascal-voc}. IoU is a score between (0,1) that effectively measures the percentage of overlap between a predicted object and a true object in a dataset. It is common to use a threshold on IoU to mark a given prediction from a localizer as a true positive (TP) or a false positive (FP). For example \cite{pascal-voc} uses 0.5 and \cite{coco} uses a range from 0.5 to 0.95 with 11 steps in between.

Given a dataset, the number of objects is the quantity $P$. A localizer then has a recall of $\frac{TP}{P}$ and a precision of $\frac{TP}{TP+FP}$. In plain language, recall gives the probability of predicting a signal to be present given that a signal is actually present. Precision gives the probability of predicting a signal to be present when there is no signal present. Both quantities are necessary because a localizer that predicts many signals can score a high recall but low precision. A localizer that never predicts a signal is present would score low recall but high precision. Precision is analogous to $1-P(fa)$ and recall is analogous to $P(D)$; however, each quantity has a distinct meaning that should not be conflated since one set of quanitities (precision and recall) is measured on a dataset with a hyperparameter (IoU) and another set of quantities ($P(fa)$ and $P(D)$) is computed using true probability distributions.

Later on, we will present the first precision and recall over SNR curves for a radiometer and compare that to a neural network based energy detector.

\section{Dataset}

In order to evaluate both the radiometer and neural network, a new dataset must be created. This dataset will also be used to train the neural network. The dataset is recorded using the SigMF format (a detached JSON header with metadata and binary files of complex int16 for data). The dataset consists of 130 SigMF records. The metadata is generated first to give 130 unique band layouts that consist of signal bursts with randomly varying modulations, bandwidths, start times, duration of bursts, and signal amplitudes.

The dataset consists of signals with the following modulation schemes:

\begin{enumerate}
    \item PSK2
    \item PSK4
    \item PSK8
    \item QAM16
    \item QAM64
    \item QAM256
    \item OFDM (always with 512 subcarriers)
    \item FSK2
    \item FSK4
    \item GMSK (FSK2 with gaussian pulse shape)
    \item OOK
    \item AM-DSB
    \item AM-SSB
    \item FM
\end{enumerate}

The digital modulations are created with random symbols and in the case of single carrier systems (other than GMSK) use a root-raised cosine pulse shape filter. The analog modulations using a variety of music and talk soundtracks pulled from youtube.com as the modulation source.

Each signal is resampled to match the bandwidth and time duration specified in the SigMF band layout. All signals are then summed to form a wideband capture with many signals present to form complete SigMF records. The raw dataset has no noise or other channel impairments other than adjacent channel interference from sidelobes and filter artifacts. This allows for the most control of SNR during training and testing since the primary objective is to establish precision and recall metrics for the signal localization task.

The dataset is partitioned in to test data and training data so that for the neural network case there is no overfitting to the test set. This dataset is available for non-commercial use at https://quadrature.dev/wideband-signal-localization-dataset/.

\section{Channelized Radiometer}

The radiometer is the most generic tool for detecting the presence of signals and as we have seem from the work on LADs it can be useful to localize signals in frequency. The radiometer used in the following discussion is a channelized radiometer operates in a block mode where the sampled bandwidth is split in to $c$ channels. Each channel is integrated for $N$ samples. A step of the radiometer computes $\Sigma_{n=0}^N |s_c(n)|^2$ where $s_c$ are the samples of a single channel. Each step of the radiometer advances $s_c$ by $N$ samples so that there are no samples in common between adjacent radiometer steps. There are $S$ steps per radiometer block. Signal localization from channelized radiometer output will occur in a post processing step that will be discussed in Section \ref{sec:radiometer-post-process}

\subsection{Radiometer Design}

The critical design choices of a channelized radiometer include 

\begin{enumerate}
    \item a noise power spectral density estimate
    \item a thresholding criteria for determining if a signal is present in a given bin or not
    \item integration length
    \item channel width
\end{enumerate}

The channel width and integration length are effectively hyperparameters that are determined by the time and frequency resolution of the the signal localizer. The thresholding criteria and noise estimate are algorithm decisions that have a major impact on the radiometer performance. In the case of a radiometer for signal localization, the post-processing routine is also a major design decision.

To determine the noise estimate, we fit a gaussian curve with the variance as a free variable to a histogram of the channelized radiometer output for bins that likely have noise without a signal. This is accomplished by computing a large number of channelized radiometer outputs and ordering the statistics by magnitude. The lower $x\%$ are considered to be noise. This results in a biased estimate; however, it is a practical bootstrapping of the radiometer that generates a noise estimate without precise calibration and allowing for signals to be present while bootstrapping the radiometer. A similar approach to estimating noise power spectral density using ordered statistics is used in \cite{lads,lad2d}.

The thresholding criteria is another important design decision. As the work on LAD shows that using multiple thresholds can catch rising and falling edges that are more robust to noise splitting a single signal in to multiple detected regions. For this study, we will use a single threshold based on the estimated variance of the gaussian fit to give a constant false alarm rate per time and frequency channel decision. 

 The following procedure is used to generate a noise variance estimate and radiometer threshold

\begin{enumerate}
	\item generate the histogram of the radiometer test statistic according to the Rice rule (number of histogram bins is $R=\frac{1}{2n}^{1/3}$ for $R$ radiometer bins and $n$ data points) \cite{rice-rule}
	\item set $k$ to the argmax (index of the max value) of the histogram
	\item store the first $2 \cdot k$ bins as $noisehistogram$
	\item use a gradient descent optimizer to minimize the mean-squared error between a scaled gaussian pulse (so that the generated window has the same height as the histogram bins for the mode) with the standard deviation being the degree of freedom for optimization
	\item use the variance of the best fit gaussian pulse to develop a threshold using a Constant False-Alarm Rate (CFAR) method
\end{enumerate}

Decisions will then be grouped together using a density based clustering technique in post-processing to get localization bounds.

\subsection{Post Processing}

\label{sec:radiometer-post-process}

The channelized radiometer output is a 2-dimensional (time and frequency) grid of statistics and binary decisions on a threshold. For the purpose of signal localization this needs to be transformed in to a bounding box that can provide time and frequency bounds. In other words, channelized radiometer decisions must be clustered in to signal decisions. Since the number of signal is not known the clustering algorithm must be capable of determining that.

Density Based Spatial Clustering with Applications to Noise (or DBSCAN) \cite{dbscan} is a popular algorithm that can simultaneously determine the number of clusters and cluster outputs; however, it has a critical flaw for this application because there is no constraint on the shape of clustering. For example, if a single bin connects two larger clusters, then DBSCAN will consider both masses to be a single cluster. This causes problems with signals that are close in time or frequency (by a small number of channel or time bins) as well as signals with deep frequency-selective fades. To keep post-processing simple, the time and frequency bounds will be assumed to be the extreme edges of a cluster, which generally means that the signal clusters should be rectangular since most communication signals are rectangular in time-frequency space. To address this, we modified DBSCAN to specifically cluster spectral regions which we call Density-based spectrogram clustering.

Density-based spectrogram clustering is initialized in the same way as DBSCAN: core points are identified by spectrogram bins (or channelized radiometer bins) that pass a magnitude threshold test with a sufficient number of neighboring bins (in $L^1$ or $L^2$). For each core point, a cluster begins by alternately expanding in each direction (forward in time, lower in frequency, backward in time, upper in frequency, etc). For each expansion, if the number of new bins that pass the radiometer decision threshold in the new region is above some threshold (for example, 50\%), then the expansion is kept. Otherwise, the expansion is rejected. The expansion continues until no direction can be expanded. If a core point becomes part of another core point's cluster then it is removed from the set of core points that need expansion. The algorithm continues for each core point until every core point belongs to a cluster that cannot be expanded.

The time bounds for each signal are the extreme bounds in the time-dimension of the cluster that forms this signal. The upper and lower frequencies are the extremes of the cluster in the frequency-dimension.

\subsection{Results}

The channelized radiometer with density-based spectrogram clustering forms a simple signal localizer. This is evaluated on the a QPSK test signal with the following parameters for the radiometer:

{
\begin{table}[htbp]
\centering
\begin{tabular}
{l|c}
Radiometer Parameter & Value \\
\hline
\hline
    Number of channels & 256 \\
    Integration length & 2 \\
    False Alarm Rate & 0.05
\end{tabular}
\end{table}
}

The test is presented to the radiometer 20000 times per SNR step with uniquely drawn AWGN. An IOU threshold of 0.5 determines whether a predicted signal is a true positive (TP) or a false positive (FP). Since the same signal is shown 20000 times per SNR step the value P for each SNR level is 20000.

\begin{figure}[!t]
\centering
\includegraphics[width=\columnwidth]{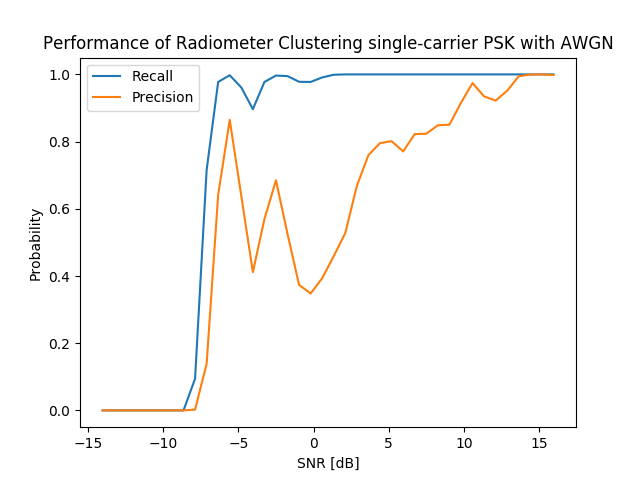}
\caption{Precision and recall for a radiometer and spectral clustering to form signal localization boundaries.}
\label{fig:radiometer-precision-recall}
\end{figure}

Figure \ref{fig:radiometer-precision-recall} shows the precision and recall over the SNR range -15 dB to 15 dB. The SNR is calculated as the total signal power over the total noise power (rather than the signal power spectral density and the in-band noise power spectral density). This is done to compare results to those presented in LAD-2D.

The recall shows a large jump from 0 to nearly 100\% around 8-9 dB SNR. The Precision shows a similar jump above 0 around 7 dB before dropping at moderate SNR and rising again as SNR increases. This is due to a large number of radiometer bins that pass the threshold and never join another cluster because they are in the roll-off region of the signal of interest. Further discussion of this phenomenon is in Section \ref{sec:discussion}

\section{Spectral Segmentation}

The entire process of the channelized radiometer can be transformed in to a well-known machine learning task used in image and video processing called segmentation. Semantic segmentation is a popular form of segmentation that classifies each pixel in an image. This is directly analogous to the radiometer task of detecting whether a time/frequency bin contains a signal or no signal. In image processing the input image is classified per pixel on the output with the same resolution as the input image.

Since we will apply a similar concept to spectral analysis where the input is time-domain samples the task will be called spectral segmentation. In order to train a deep neural network for spectral segmentation, the following choices must be made

\begin{itemize}
    \item loss function
    \item network architecture
    \item frequency resolution
    \item time resolution
\end{itemize}

The loss function for a balanced segmentation dataset for signal detection can be a simple binary cross-entropy to decide if a given cell in the time-frequency grid contains a signal or no signal.

U-net \cite{unet} is a popular choice for segmentation tasks due to its ability to gather features at multiple scales with minimal distortion in the upsampling process (as compared to SegNet) which has set performance benchmarks on challenging medical imagery tasks such as \cite{isbi-challenges}.

For the results in this paper a frequency resolution of 512 bins (on the synthetic dataset). Since no overlap is taken between rows, there will be a 512-sample time resolution.

\subsection{Neural Network Design}

 Since the input to spectral segmentation is actually time-domain complex baseband samples, u-net requires some transformation from the 1-d representation to a 2-d representation with the same dimensions as the desired time/frequency grid on the output of the spectral segmentation task. Many transformations are possible; however, for the purpose of establishing a baseline on the task we will use a normalized log-magnitude spectrogram. For a frequency resolution of 512 bins the input samples are taken in chunks of 512 samples with no overlap, and an absolute value of the Discrete Fourier Transform (DFT) gives a spectrogram. The log of this spectrogram is then normalized by removing the mean and normalizing the magnitude by the standard deviation of the spectrogram.

\subsubsection{Training}

The previously described dataset consists of 260 training files. Each file contains 100 million samples with a random and unique band layout forming 12425 unique signals across those files. Training uses the Adam optimizer \cite{adam-optimizer} with a learning rate of 3e-4. Each epoch consists of 25 training steps followed by 25 validation steps with the average loss across each of the training and validation steps (respectively) recorded.

Training data has AWGN added with a random standard deviation uniformly distributed between 1e-9 to 1e-4. This gives an SNR range of 30 dB to -10 dB for this dataset. The validation data is randomly drawn from the training set, but with AWGN added using a constant standard deviation of 1e-5.

\begin{figure}[!t]
\centering
\includegraphics[width=\columnwidth]{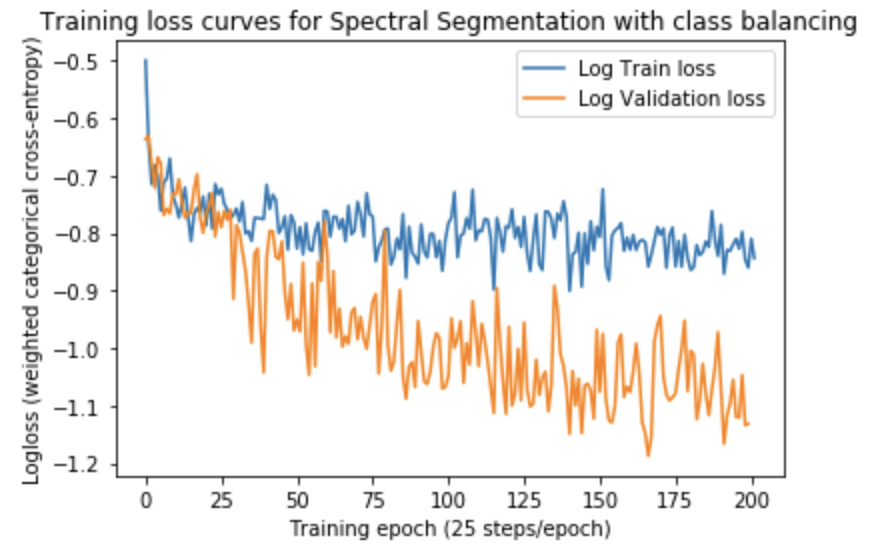}
\caption{Training loss for the neural network spectral segmentation training.}
\label{fig:nn-training-loss}
\end{figure}

The network is trained for 200 epochs with the log value of the training and loss curves shown in Figured \ref{fig:nn-training-loss}. The notable feature of this curve is that the training loss does not appear to change after the 50th epoch; however, the lower SNR range of the validation loss continues to improve indicating that learning the higher SNR cases is much easier for the network and further training improves the lower SNR response.

\subsection{Post Processing}

The forward pass of the trained spectral segmentation network results in semantically similar output of the channelized radiometer. However, as will be shown in the following section on results the post-processing algorithm can be relaxed due to improved detection and threshold accomplished inside the neural network. Instead of the more complex density-based spectrogram clustering, the spectral segmentation network can be processed using a standard connected components labeling algorithm which has many well optimized variations. Connected components labels each connected region as a unique cluster. The extreme bounds of each cluster are assumed to be the time and frequency bounds of the detected signal.

\subsection{Results}

\begin{figure}[!t]
\centering
\includegraphics[width=\columnwidth]{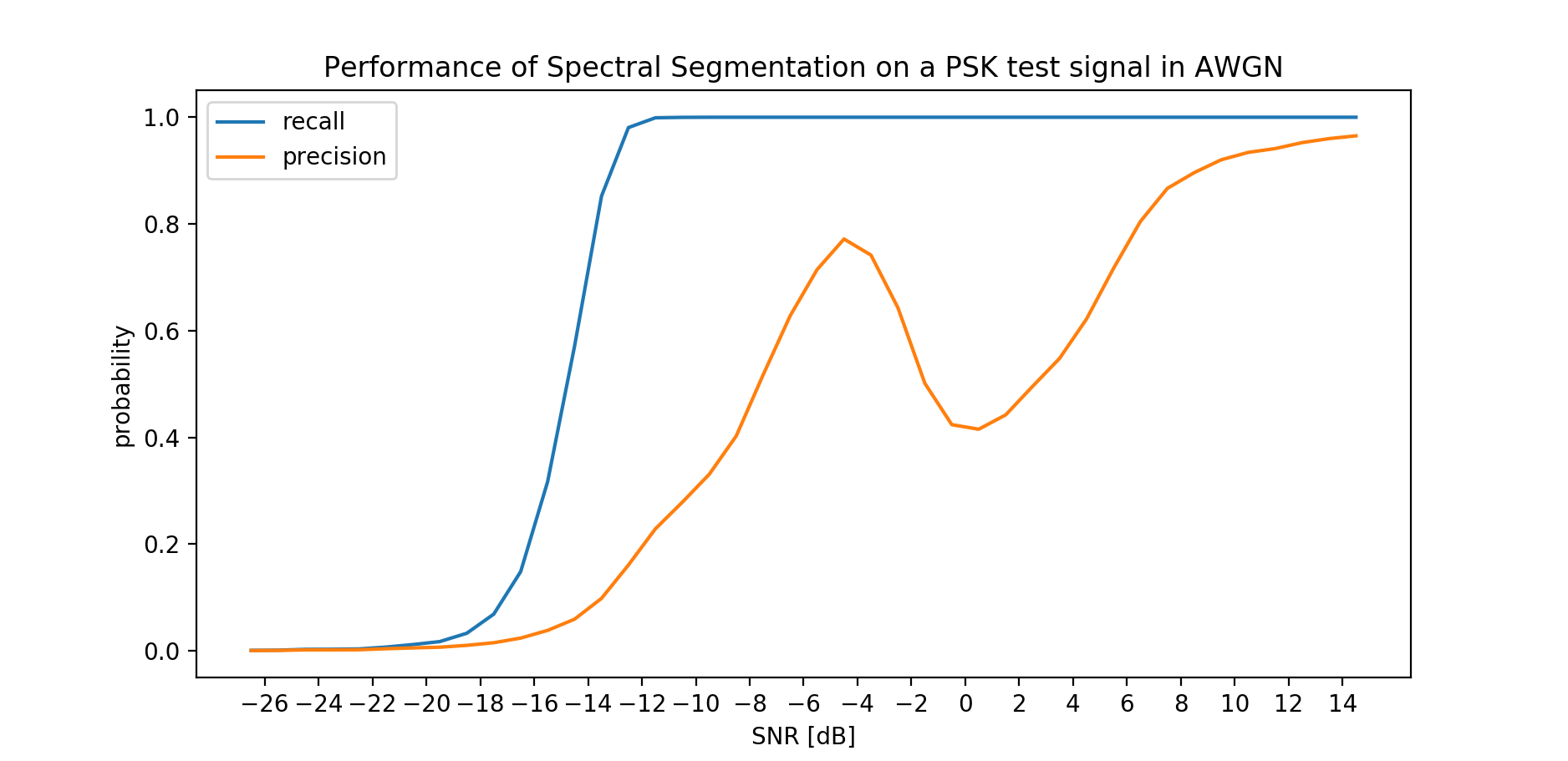}
\caption{Precision and recall for a neural network performing spectral segmentation followed by connected components to form signal localization boundaries.}
\label{fig:nn-precision-recall}
\end{figure}

Figure \ref{fig:nn-precision-recall} shows the precision and recall for the spectral segmentation u-net with connected components post-processing over a range of SNRs with AWGN. For each SNR step the same test file with 100 million samples of QPSK with 5x oversampling is used with uniquely drawn AWGN is looped over 20 times for a total of 30517 unique test vectors. The recall shows at 8 dB improvement over the channelized radiometer with a single threshold and more complex post-processing with no change in precision.

The precision over SNR follows a similar pattern to the channelized radiometer from Figure \ref{fig:radiometer-precision-recall}. The initial peak is higher than the radiometer and the valley is not as low. Eventually the precision approaches 1.0 as SNR increases. The following section contains a discussion on this phenomenon and how to deal with it.

\section{Discussion}
\label{sec:discussion}

The precision and recall curves scored on true test data in AWGN show that a neural network trained for spectral segmentation makes a superior signal detector with simple connected components labeling post-processing than a channelized radiometer with careful hyperparameter selection and more complex density-based clustering. Both detectors exhibit a sharp rise in recall at some SNR where signals become detectable and stay near 100\% above that SNR. For the neural network this point appears 8 dB before it does on the same test data for the channelized radiometer. The results from the radiometer match well with what would be expected for a recall score from similar work in \cite{lad2d}. The LAD-2d algorithm from \cite{lad2d} was not tested at very low SNRs, so it is unclear if they would exhibit a similar sharp rise in detection; however, since LAD-2d exhibits an increase from 90\% to 100\% between -7 and -8 dB the performance would be on par with the presented channelized radiometer.

\begin{figure}[!t]
\centering
\includegraphics[width=\columnwidth]{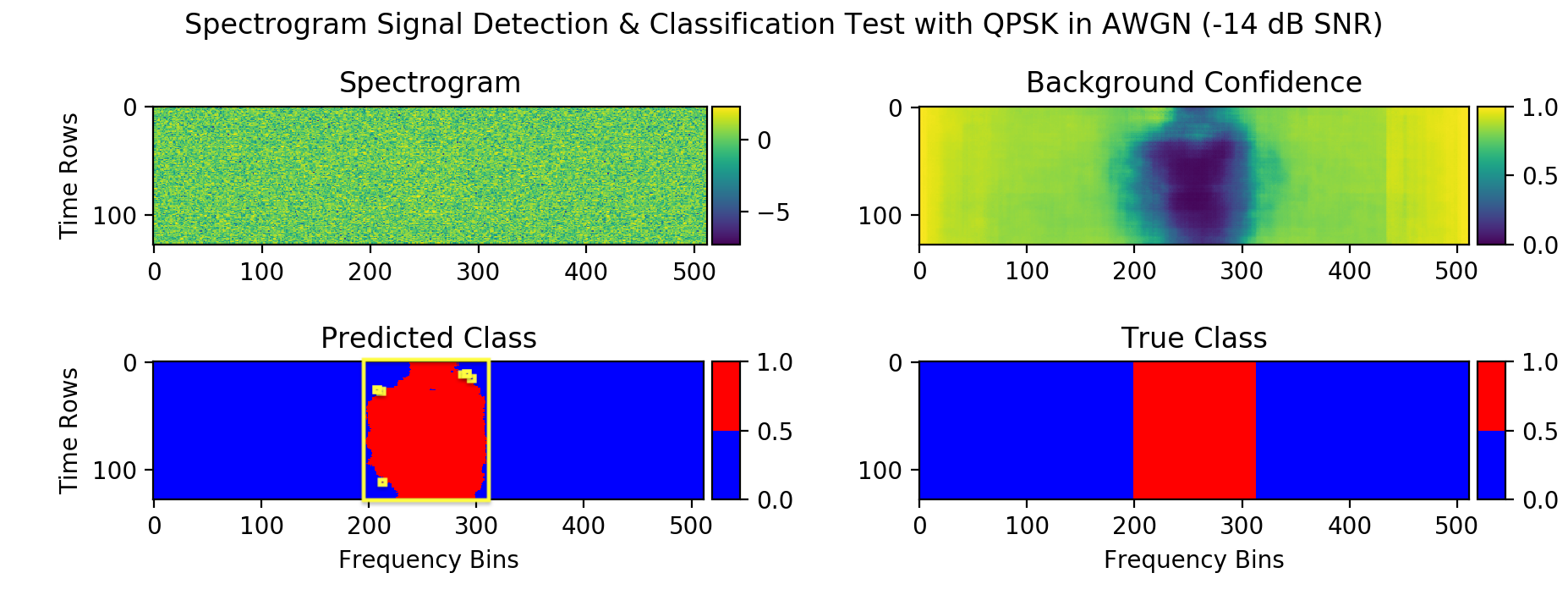}
\caption{A quad-chart showing a spectrogram of the test example at top left, the network raw output at top right, thresholded at bottom left, and the true (expected) output at bottom right. This example is -14 dB of SNR and the signal is only visible with long averaging. The network is able to detect most of the signal region as having a signal present with several disjoint detections that causes a low precision score.}
\label{fig:low-precision-nn}
\end{figure}

\begin{figure}[!t]
\centering
\includegraphics[width=\columnwidth]{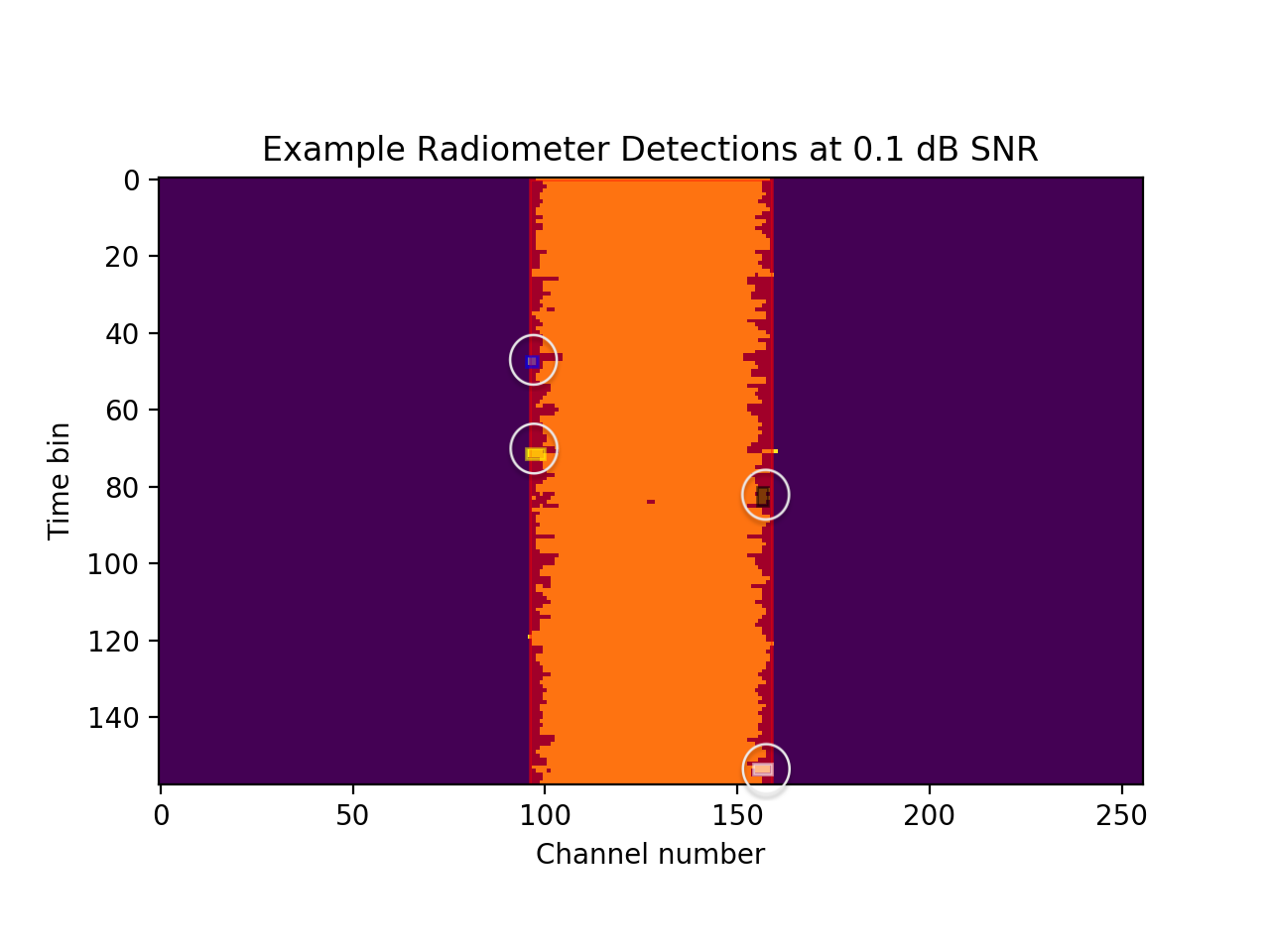}
\caption{A 0 dB SNR example from the channelized radiometer. The regions that pass the radiometer threshold at shown in yellow overload with the true region in a transparent orange. The regions that are detected as signal, but disjoint from the primary signal detection are circled. These regions cause a low precision score for this example.}
\label{fig:low-precision-radiometer}
\end{figure}

An interesting phenomenon for both detectors is an initial increase in precision followed by a valley at moderate SNRs before eventually climbing. This is valley is caused by very small signal regions being detected as signals that are disjoint from the primary signal detection. Figure \ref{fig:low-precision-nn} shows an quad-chart example from the neural network that shows this happening. The example shows at -14 dB SNR signal that is detected by the network with a single component that overlaps well with the true signal region. However, there are 5 different regions within this example of 1-2 bins (that are within the true signal boundary) that do not get connected to the larger detected mass. This example would have a precision of 1/6. A similar example with moderate SNR is shown for the radiometer in Figured \ref{fig:low-precision-radiometer}. This example only shows the radiometer detection mask overlaid with a transparent mask of the true signal region. In this case, the primary signal region detects a signal with 4 very small regions detecting a signal, but disjoint from the primary signal detection. Low SNR examples exhibit an initial increase due to the primary signal region being detected at all. Moderate SNRs show an increase in these single bin detections within the roll-off region of a signal.

This decrease in precision is effectively a sharp rise in false detections at moderate SNRs. These can be dealt with in a number of ways. The radiometer has greater flexibility to use more complex thresholding parameters such as those presented in \cite{lad2d} to reduce these single-bin detections. Both methods would benefit from filtering abnormally small (such as 1-bin) detections which introduces another heuristic with knowledge of the expected signals. A final approach would be the gather all detections that are entirely contained within another region as a single signal. This would dramatically improve precision scores over the entire range of SNR values and especially on the fringe edges of true signal regions such as those in the moderate SNR regions.

The downside to this approach to signal detection remains detecting instances of signals that occur very close in time and frequency. The training and test set have not tested performance in fading channels; however, it is expected and known that a radiometer will perform very poorly in fading environments.

\section{Conclusion}

The signal localization problem for wideband spectrum sensing requires detecting potentially multiple signals within the sample bandwidth at arbitrary center frequencies, offsets, and time bounds. Algorithms that solve this problem can be compared well using precision and recall with an IoU threshold as the figures of merit. A channelized radiometer is developed with a new density-based clustering as a post-processor to solve this problem with precision and recall presented. This compares favorably with the only known existing work on blind signal localization using similar test data and procedure. A novel neural network training regime is introduced applying segmentation as a signal localizer with connected components as the post-processing to transform network output in to signal localization predictions. This improves the channelized radiometer recall by 8 dB with similar precision. The poor precision results in both approaches are explained with examples and a path to approach the problem in future work.








\bibliographystyle{IEEEtran}
\bibliography{IEEEabrv,phd-bib.bib}

\begin{thebibliography}{10}
\providecommand{\url}[1]{#1}
\csname url@samestyle\endcsname
\providecommand{\newblock}{\relax}
\providecommand{\bibinfo}[2]{#2}
\providecommand{\BIBentrySTDinterwordspacing}{\spaceskip=0pt\relax}
\providecommand{\BIBentryALTinterwordstretchfactor}{4}
\providecommand{\BIBentryALTinterwordspacing}{\spaceskip=\fontdimen2\font plus
\BIBentryALTinterwordstretchfactor\fontdimen3\font minus
  \fontdimen4\font\relax}
\providecommand{\BIBforeignlanguage}[2]{{%
\expandafter\ifx\csname l@#1\endcsname\relax
\typeout{** WARNING: IEEEtran.bst: No hyphenation pattern has been}%
\typeout{** loaded for the language `#1'. Using the pattern for}%
\typeout{** the default language instead.}%
\else
\language=\csname l@#1\endcsname
\fi
#2}}
\providecommand{\BIBdecl}{\relax}
\BIBdecl

\bibitem{snr-walls}
R.~Tandra and A.~Sahai, ``Snr walls for signal detection,'' \emph{IEEE Journal
  of Selected Topics in Signal Processing}, vol.~2, no.~1, pp. 4--17, Feb 2008.

\bibitem{wb-nb-detection}
J.~Lehtomaki, J.~Vartiainen, and M.~Juntti, ``Combined wideband and narrowband
  signal detection for spectrum sensing,'' in \emph{2009 Second International
  Workshop on Cognitive Radio and Advanced Spectrum Management}, May 2009, pp.
  91--95.

\bibitem{snr-walls-features}
R.~Tandra and A.~Sahai, ``Snr walls for feature detectors,'' in \emph{2007 2nd
  IEEE International Symposium on New Frontiers in Dynamic Spectrum Access
  Networks}, April 2007, pp. 559--570.

\bibitem{energy-detection}
H.~Urkowitz, ``Energy detection of unknown deterministic signals,''
  \emph{Proceedings of the IEEE}, vol.~55, no.~4, pp. 523--531, April 1967.

\bibitem{adaptive-threshold}
A.~Gorcin, K.~A. Qaraqe, H.~Celebi, and H.~Arslan, ``An adaptive threshold
  method for spectrum sensing in multi-channel cognitive radio networks,'' in
  \emph{2010 17th International Conference on Telecommunications}, April 2010,
  pp. 425--429.

\bibitem{xgcomms}
M.~P. Olivieri, G.~Barnett, A.~Lackpour, A.~Davis, and P.~Ngo, ``A scalable
  dynamic spectrum allocation system with interference mitigation for teams of
  spectrally agile software defined radios,'' in \emph{First IEEE International
  Symposium on New Frontiers in Dynamic Spectrum Access Networks, 2005. DySPAN
  2005.}, Nov 2005, pp. 170--179.

\bibitem{survey-of-spectrum-sensing}
T.~Yucek and H.~Arslan, ``A survey of spectrum sensing algorithms for cognitive
  radio applications,'' \emph{IEEE Communications Surveys Tutorials}, vol.~11,
  no.~1, pp. 116--130, First 2009.

\bibitem{lads}
J.~Vartiainen, J.~J. Lehtomaki, and H.~Saarnisaari, ``Double-threshold based
  narrowband signal extraction,'' in \emph{2005 IEEE 61st Vehicular Technology
  Conference}, vol.~2, May 2005, pp. 1288--1292 Vol. 2.

\bibitem{lad-acc}
J.~Vartiainen, H.~Sarvanko, J.~Lehtomaki, M.~Juntti, and M.~Latva-aho,
  ``Spectrum sensing with lad-based methods,'' in \emph{2007 IEEE 18th
  International Symposium on Personal, Indoor and Mobile Radio Communications},
  Sept 2007, pp. 1--5.

\bibitem{localization-fcme}
J.~Vartiainen, J.~J. Lehtomäki, S.~Aromaa, and H.~Saarnisaari, ``Localization
  of multiple narrowband signals based on the fcme algorithm,'' in \emph{Nordic
  Radio Symposium}, aug 2004.

\bibitem{lad2d}
J.~Vartiainen, J.~LEHTOMAeKI, H.~Saarnisaari, M.~Juntti, and K.~Umebayashi,
  ``Two-dimensional signal localization algorithm for spectrum sensing,''
  \emph{IEICE transactions on communications}, vol.~93, no.~11, pp. 3129--3136,
  2010.

\bibitem{mathys-multi-sc}
\BIBentryALTinterwordspacing
P.~Mathys, ``Efficient band occupancy and modulation parameter detection,''
  \emph{Proceedings of the GNU Radio Conference}, vol.~2, no.~1, p.~8, 2017.
  [Online]. Available:
  \url{https://pubs.gnuradio.org/index.php/grcon/article/view/36}
\BIBentrySTDinterwordspacing

\bibitem{pascal-voc}
M.~Everingham, L.~Van~Gool, C.~K.~I. Williams, J.~Winn, and A.~Zisserman, ``The
  pascal visual object classes (voc) challenge,'' \emph{International Journal
  of Computer Vision}, vol.~88, no.~2, pp. 303--338, Jun. 2010.

\bibitem{coco}
\BIBentryALTinterwordspacing
T.~Lin, M.~Maire, S.~J. Belongie, L.~D. Bourdev, R.~B. Girshick, J.~Hays,
  P.~Perona, D.~Ramanan, P.~Doll{\'{a}}r, and C.~L. Zitnick, ``Microsoft
  {COCO:} common objects in context,'' \emph{CoRR}, vol. abs/1405.0312, 2014.
  [Online]. Available: \url{http://arxiv.org/abs/1405.0312}
\BIBentrySTDinterwordspacing

\bibitem{rice-rule}
D.~W. Scott, ``\BIBforeignlanguage{eng}{Histogram},''
  \emph{\BIBforeignlanguage{eng}{Wiley Interdisciplinary Reviews: Computational
  Statistics}}, vol.~2, no.~1, pp. 44--48, 2010.

\bibitem{dbscan}
M.~Ester, H.-P. Kriegel, J.~Sander, X.~Xu \emph{et~al.}, ``A density-based
  algorithm for discovering clusters in large spatial databases with noise.''
  in \emph{Proceedings of the Second International Conference on Knowledge
  Discovery and Data Mining}, August 1996.

\bibitem{unet}
O.~Ronneberger, P.~Fischer, and T.~Brox, ``U-net: Convolutional networks for
  biomedical image segmentation,'' in \emph{Medical Image Computing and
  Computer-Assisted Intervention -- MICCAI 2015}, N.~Navab, J.~Hornegger, W.~M.
  Wells, and A.~F. Frangi, Eds.\hskip 1em plus 0.5em minus 0.4em\relax Cham:
  Springer International Publishing, 2015, pp. 234--241.

\bibitem{isbi-challenges}
\BIBentryALTinterwordspacing
S.~K. Gehlot and A.~Gupta, ``Isbi challenge workshops,'' in \emph{International
  Symposium on Biomedical Imaging}, April 2019. [Online]. Available:
  \url{https://biomedicalimaging.org/2019/challenges/}
\BIBentrySTDinterwordspacing

\bibitem{adam-optimizer}
\BIBentryALTinterwordspacing
D.~P. Kingma and J.~Ba, ``Adam: {A} method for stochastic optimization,''
  \emph{CoRR}, vol. abs/1412.6980, 2014. [Online]. Available:
  \url{http://arxiv.org/abs/1412.6980}
\BIBentrySTDinterwordspacing

\end{thebibliography}
%



%

\begin{IEEEbiography}[{\includegraphics[width=1in,height=1.25in,clip,keepaspectratio]{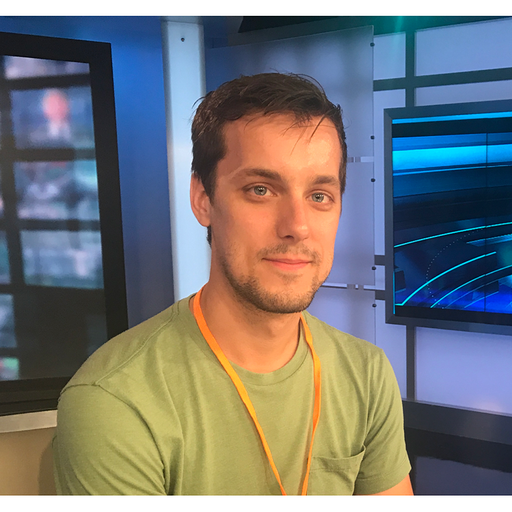}}]{Nathan West}
Nathan West is pursuing a PhD at Oklahoma State University in the area of wideband spectrum sensing and pursuing machine learned methods. He is also the Director of Machine Learning at DeepSig. At DeepSig he is interested in high-rate software defined radio using machine learning inspired methods to derive approximations to communications signal processing challenges that can yield greater accuracy and lower computational complexity than traditional methods. He believes strongly in reproducible machine learning for the wireless communications field by sharing public datasets and code along with publications. Previously, he has been a core contributor to GNU Radio and maintained the VOLK project for several years.
\end{IEEEbiography}

\begin{IEEEbiographynophoto}{Tamoghna Roy}
Tamoghna Roy is the principal Machine Learning engineer at DeepSig and has a PhD from Virginia Tech University. He is an expert in non-wiener characteristics of adaptive LMS equalizers and has published extensively on machine learning-based methods of approximating wireless channels.
\end{IEEEbiographynophoto}


\begin{IEEEbiographynophoto}{Tim O'Shea}
Tim O'Shea is the CTO at DeepSig and associate professor at the Virginia Tech Hume Center. He has pioneered the field of machine learning for communications systems since his PhD at Virginia Tech University. He has published extensively on machine-learned methods for communication systems and is a frequent invited speaker. He has contributed to open source software defined radio through GNU Radio and a variety of other projects.
\end{IEEEbiographynophoto}


\vfill


\end{document}